\documentclass[aps,prl,twocolumn,showpacs]{revtex4} 
\usepackage{bm,amssymb,stmaryrd}

\newcommand{\sgn}{\mathop{\mathrm{sgn}}\nolimits}  
  
\newcommand{\STr}{\mathop{\mathrm{STr}}\nolimits}  
\renewcommand{\vec}[1]{\bm{#1}}

\begin{document}

\title{Quantum Anomaly and Effective Field Description of a Quantum
  Chaotic Billiard}
 
\author{Nobuhiko Taniguchi}

\affiliation{Institute of Physics, University of Tsukuba, Tennodai Tsukuba
  305-8571, Japan}

\date{\today}

\begin{abstract}
We investigate the effective field theory of a quantum chaotic
billiard from a new perspective of quantum anomalies, which result
from the absence of continuous spectral symmetry in quantized
systems.  It is shown that commutators of composite operators on the
energy shell acquire anomalous part.  The presence of the anomaly
allows one to introduce effective dual fields as phase variables
without any additional coarse-graining nor ensemble averaging in a
ballistic system.  The spectral Husimi function plays a role as the
corresponding amplitude.
\end{abstract}%


\maketitle%


The study of quantum chaos, namely, quantum properties of classically
nonintegrable systems, has been attracting much attention over two
decades.  By the presence of irregular boundary and/or impurities
inside, a quantum billiard system falls into this category, and it has
been a rich source of research not only as a manageable model bridging
between chaos and quantum mechanics but also as a model of electronic
devices such as quantum dots.
With the standard quantization method heavily relying on the presence
of invariant tori, the quantization prescription of classically
chaotic systems is not yet fully disclosed.  Yet the success of
quantum mechanics is so striking that it seems sensible to seek what is
a quantum signature of classically chaotic systems by assuming the
validity of the Schr\"{o}dinger equation.

The symmetry of quantum theories may be substantially different from
the classical one since it is not always possible to retain all the
classical symmetries at quantum level.  When some classical symmetry
is broken, one may call a quantum system anomalous and detect it by
the presence of quantum anomalies \cite{Anomaly}.  Often, quantum
anomalies will matter in the context of gauge field theories, yet the
notion goes far beyond it and serves as a fundamental feature of
quantum theories.

In this Letter, we investigate the effective field description of a
quantum chaotic billiard from a novel perspective --- quantum
anomalies in spectra.  The relevance of the anomaly to a quantum
billiard is most easily understood by noting different spectral
structures between classical and quantum theories.  Whereas classical
dynamics has continuous symmetry along the energy  by changing
the momentum continuously without altering the orbit in space, such
continuous symmetry is absent quantum mechanically since discrete
energy levels are formed.  In this sense, the anomaly here is
taken as a revelation of the quantization condition.
By examining the algebraic structure of the ``current algebra'', we
will show the presence of anomalous part (Schwinger term) in current
commutators.  Accordingly, it enables us naturally to construct
effective fields as phase variables \emph{without} any additional
coarse-graining nor ensemble averaging, whereas the spectral Husimi
function shows up and acts as the amplitude (see
Eqs.~(\ref{eq:effective1},\ref{eq:effective2}) below).

Effective field theories by the supermatrix nonlinear-sigma
(NL-$\sigma$) model have been quite successful in describing
disordered metals with diffusive dynamics \cite{EfetovBook}.  In
addition to explaining weak localization phenomena, the zero-mode
approximation has provided a \emph{direct proof} of the
Bohigas-Giannoni-Schmit conjecture stating that the level
correlations of quantum chaotic systems in general obey the
Wigner-Dyson statistics from random matrices \cite{Bohigas84}.
By seeing the success of effective field description of diffusive
chaotic systems, the same zero-dimensional model has been anticipated
by the universality in a ballistic system without any intrinsic
stochasticity nor disorder, and a great effort has been put
forth to extend the framework toward it.
Over last decade, several ``derivations'' of ballistic NL-$\sigma$
models, which is claimed to be applicable to the length scale shorter
than the mean free path, have been proposed, based on a quasiclassical
approximation \cite{Muzykantskii95Efetov03}, on the ensemble averaging
either over energy spectra \cite{Andreev96,Altland99b} or over the
external parameter \cite{Zirnbauer99b}, or by a functional
bosonization approach \cite{Efetov04}.  Meanwhile our understanding
has progressed considerably in particular on the connection between
the statistical quantum properties and the classical chaotic dynamics.

The validity of these ballistic NL-$\sigma$ models, however, is not so
transparent unlike the diffusive counterpart.  Soon after the
derivations, it has been recognized that some (unphysical) zero mode
exists along a vertical direction of the energy shell and nothing
suppresses those fluctuations \cite{Altland99b,Zirnbauer99b}.  As a
result, how to attain the necessary ``mode-locking'' has been
disputed.  Though similar difficulty is absent in some approaches
\cite{Muzykantskii95Efetov03,Efetov04}, the mode-locking mechanism in
those works is ascribed to assuming the existence of the Fermi
surface.  This is rather odd because the Fermi surface is a many-body
effect and has nothing to do with the notion of quantum chaos.
In the present work, we will find an explanation by the anomaly
carried by each level.  The mechanism is of one-body nature and 
applicable either to noninteracting bosons or noninteracting
fermions.  

In general, quantum anomalies can be detected in current commutators
by anomalous part (Schwinger term) proportional to the derivative of
the Dirac delta function.  There are several ways to reveal such
contribution \cite{Anomaly}: point-splitting methods, normal ordering
prescription, cohomological consideration, examining the functional
Jacobian etc.  Among them, we choose the normal ordering prescription
with the canonical quantization scheme (see also \cite{Isler88Plus}).
In contrast to previous effective field theories with functional
integration, we find that the present approach clarifies the subtlety
hidden in regularization in a clearer and more pedagogical way.

The \emph{intrinsic} need of regularization stems from the presence of
discretized energy levels $\varepsilon_{\alpha}$.  In energy
integration, the effect is accommodated by an insertion of the Dirac
delta function $\delta(\varepsilon - \varepsilon_{\alpha})$, but its
singular nature requires some regularization to make the theory
finite.  A standard way is to define the Dirac delta function with
positive infinitesimal $\eta$ by
\begin{equation}
  \delta(\varepsilon-\varepsilon_{\alpha})= \frac{i}{2\pi} \left(
  \frac{1}{\varepsilon - 
  \varepsilon_{\alpha} + i\eta} - \frac{1}{\varepsilon -
  \varepsilon_{\alpha} - i\eta} \right). 
\label{eq:smooth-delta}
\end{equation}
A crucial observation in the present context is to view it as an
embodying the point-splitting method along the energy axis, hence a
field theory defined on the energy coordinate is called for.  The
point-splitting regularization is known to be equivalent to
introducing the normal-ordered operator, so that we proceed it by
defining the appropriate vacuum $|0\rangle$ and creation/annihilation
parts of operators.


Having in mind the Hamiltonian $\mathcal{H}$ describing quantum
dynamics in an irregular confinement potential, we begin with the
Schr\"{o}dinger equation in the first quantized form, $\mathcal{H}
\phi_{\alpha}(\vec{r}) = \varepsilon_{\alpha} \phi_{\alpha}(\vec{r}) $
with eigen energies $\varepsilon_{\alpha}$ and eigen functions
$\phi_{\alpha}(\vec{r})$.  We consider a \emph{generic} quantum
chaotic billiard, by which we mean that neither spectral degeneracy
nor dynamical symmetry intertwining the spectra exists.  In this
situation, it makes sense to attach independent creation/annihilation
operators $\psi^{\dagger}_{\alpha}$ and $\psi_{\alpha}$ to each level
$\alpha$.  Though operators $\psi_{\alpha}$ may be either bosonic or
fermionic, we first assume them as fermionic, which highlights a
similarity of the present construction to the conventional
bosonization in one-dimensional electrons.  By these preparations, we
define the field operator $\psi(\vec{r}) = \sum_{\alpha}
\phi_{\alpha}(\vec{r}) \psi_{\alpha}$ obeying the commutation relation
$\{ \psi(\vec{r}_{1}),\psi^{\dagger}(\vec{r}_{2})\} =
\delta(\vec{r}_{1} - \vec{r}_{2})$.  Since we work on a noninteracting
system, we can immediately write down the commutation relation not
only at equal time but also at different time.  When we further
introduce the field operator on a certain energy shell $\varepsilon$
by
\begin{equation}
 \psi (\vec{r}t) = \int^{\infty}_{-\infty}
 \psi(\vec{r}\varepsilon) \, e^{-\frac{i}{\hbar} \varepsilon
 t}d\varepsilon, 
\end{equation}
they are found to satisfy the commutation relation
\begin{eqnarray}
&& \left\{ \psi (\vec{r}_{1} \varepsilon_{1}), \psi^{\dagger}
  (\vec{r}_{2} \varepsilon_{2} ) \right\}.
\nonumber \\ && \qquad 
= \delta(\varepsilon_{1}-\varepsilon_{2})\;\langle \vec{r}_{1}
  |\,\delta(\varepsilon_{2}-H) \,| \vec{r}_{2}\rangle\, ,
\label{eq:spectral-comm}
\end{eqnarray}
where $\delta(\varepsilon - H ) = \sum_{\alpha} |\alpha\rangle
\delta(\varepsilon - \varepsilon_{\alpha}) \langle \alpha |$ is the
spectral operator.  The above shows an intriguing feature as a field
theory that field operators on the energy shell are nonlocal in space
because of the nonlocality of the spectral operator.  Simultaneously,
apart from nonlocality, the system may be viewed as the
one-dimensional system along the ``$\varepsilon$-axis'' with some
internal degrees of freedom attached.  We will pursue this line of
description below.

The field $\psi(\vec{r}\varepsilon)$ is a subtle object because it
exists when $\varepsilon$ coincides with one of eigen energy levels
$\varepsilon_{\alpha}$ and divergence occurs due to the Dirac delta
function.  The latter divergence is regularized by Eq.
(\ref{eq:smooth-delta}) but the prescription is incomplete for
composite fields.  The nonlocal commutation relation requires us to
consider a nonlocal current operator (or density operator in usual
terminology), and we need to define it by normal-ordering by
decomposing operators into $(\pm)$ parts, $\psi = \psi_{+} + \psi_{-}$
\cite{Isler88Plus}.  Explicitly, 
projected field operators $\psi_{\pm}$ are defined by
\begin{equation}
 \psi_{\pm}(\vec{r}\varepsilon) = \frac{\pm i}{2\pi}
  \int^{\infty}_{-\infty} 
  \frac{\psi(\vec{r}\varepsilon')}{\varepsilon-\varepsilon'\pm
  i\eta}\, d\varepsilon', 
\label{eq:projectedfield}
\end{equation}
and $\psi_{\pm}^{\dagger}$ by its hermitian conjugate.  Subsequently
we introduce the vacuum state (the Dirac sea) $|0\rangle$ by requiring
$\psi_{+} |0 \rangle = \psi^{\dagger}_{-} |0 \rangle = 0$.  It means
that $\psi_{+}$ and $\psi_{-}^{\dagger}$ ($\psi_{+}^{\dagger}$ and
$\psi_{-}$) are annihilation (creation) operators and the
normal-ordered products $: \: :$ are defined accordingly.  The
commutation relation Eq. (\ref{eq:spectral-comm}) is modified for the
projected fields to be
\begin{eqnarray}
&& \big\{ \psi_{\pm} (\vec{r}_{1}, \varepsilon_{1}),
  \psi^{\dagger}_{\pm} (\vec{r}_{2}, \varepsilon_{2})
  \big\} 
\nonumber \\ &&\quad 
= \frac{(\pm i/2\pi)}{\varepsilon_{1}-\varepsilon_{2} \pm i\eta} \,
\langle \vec{r}_{1}| \delta(\varepsilon_{2}-H) | \vec{r}_{2}  \rangle,
\label{eq:spectral-comm2}
\end{eqnarray}
and all the other commutation relations vanish.  By using this
prescription, the normal-ordered current operator is defined by
\begin{equation}
  j(\vec{r},\vec{r}';\varepsilon) = \;
  :\psi^{\dagger}(\vec{r}'\varepsilon) \psi(\vec{r}\varepsilon) : \, .
\end{equation}

To see how anomalous contribution emerges in the current commutator,
it suffices to examine the vacuum average, which can be evaluated
 by the help of Eq. (\ref{eq:spectral-comm2}) and
$2\pi \delta'(z) = -(z+i\eta)^{-2}+(z-i\eta)^{-2}$ as
\begin{eqnarray}
&& \langle 0 | \left[ j(\vec{r}_{1},\vec{r}'_{1};\varepsilon_{1}),
   j(\vec{r}_{2},\vec{r}'_{2}; \varepsilon_{2}) \right]
   |0\rangle \nonumber \\ 
&&\quad =  \frac{i}{2\pi} \langle \vec{r}_{2} |
   \delta(\varepsilon_{1}-H) |\vec{r}'_{1} \rangle 
   \langle \vec{r}_{1} | \delta(\varepsilon_{2} - H) |\vec{r}'_{2} \rangle \,
 \delta'(\varepsilon_{1} -  \varepsilon_{2}). \nonumber \\
\end{eqnarray}
The right-hand side signifies anomalous contribution, {\it i.e.},
Schwinger term.  It is present only when the spatial correlation of
the spectral operator exists.
Having identified the anomalous part, we can immediately restore the
current algebra as
\begin{eqnarray}
&& \left[ j(\vec{r}_{1},\vec{r}'_{1};\varepsilon_{1}),
   j(\vec{r}_{2},\vec{r}'_{2}; \varepsilon_{2}) \right] \nonumber \\
&&\quad  = \delta(\varepsilon_{1} - \varepsilon_{2}) \left[ \langle
   \vec{r}_{1}|\delta(\varepsilon_{2} - H) |\vec{r}'_{2} \rangle
   j(\vec{r}'_{1},\vec{r}_{2}; \varepsilon_{2}) - (1\leftrightarrow 2)
   \right]\nonumber \\ 
&&\qquad +  \frac{i}{2\pi} \langle \vec{r}_{2} |
   \delta(\varepsilon_{1}-H) |\vec{r}'_{1} \rangle 
   \langle \vec{r}_{1} | \delta(\varepsilon_{2} - H) |\vec{r}'_{2} \rangle 
 \delta'(\varepsilon_{1} -  \varepsilon_{2}). \nonumber \\
\label{eq:current-comm1}
\end{eqnarray}

The above determines the current algebra on the energy shell
completely, but working on an bilocal operator of the form
$\mathcal{O}(\vec{r},\vec{r}')$ is not so convenient.  A way to
circumvent the difficulty is to recast it into an object defined on
the classical phase space $\vec{x}=(\vec{q},\vec{p})$ by taking it as
$\mathcal{O}(\vec{r},\vec{r}') = \langle
\vec{r}|\hat{\mathcal{O}}|\vec{r}'\rangle$ (still an operator).
In previous approaches
\cite{Muzykantskii95Efetov03,Andreev96,Altland99b,Zirnbauer99b,Efetov04},
the Wigner-Weyl representation has been widely utilized with a
semiclassical approximation.  Nevertheless, we find that the use of
the Husimi representation (the wave-packet representation) not only
has some advantages but also mandatory to identify the exact symmetry
of the algebra.

The Husimi representation of the operator $\hat{\mathcal{O}}$ is
defined by $\mathcal{O}(\vec{x}) = \langle \vec{x}|
\hat{\mathcal{O}}\, |\vec{x} \rangle$ where the coherent state
$|\vec{x} \rangle$ centered at $\vec{x}=(\vec{q},\vec{p})$ is defined
by
\begin{equation}
  \langle \vec{r}|\vec{x}\rangle =(\pi \hbar)^{-\frac{d}{4}}\, 
  e^{-\frac{1}{2 \hbar}(\vec{r}-\vec{q})^{2} + \frac{i}{\hbar}
  \vec{p}\cdot (\vec{r}-\frac{1}{2}\vec{q})}.
\end{equation}
Note that the coherent basis $|\vec{x}\rangle$ is overcomplete, so
that one can determine the operator uniquely by its diagonal element
while it is never the case in ordinary complete bases.  
Following this convention, we identify the Husimi representation of
$j$ as $\langle \vec{x}|j(\varepsilon) |\vec{x}\rangle$.  The
definition can be equally rewritten in terms of the field operator
$\psi(\vec{x}) = \int \!\! d\vec{r} \langle \vec{x}|\vec{r}\rangle
\psi(\vec{r})$ annihilating the wave-packet centered at
$\vec{x}=(\vec{q},\vec{p})$ as
\begin{eqnarray}
&&  j(\vec{x};\varepsilon) = \; :\psi^{\dagger}(\vec{x};\varepsilon)
  \psi(\vec{x};\varepsilon):\,.
\end{eqnarray}
It is checked that the operator $\psi(\vec{x};\varepsilon)$ obeys the
commutation relation $\{ \psi(\vec{x}; \varepsilon_{1}),
\psi^{\dagger} (\vec{x}; \varepsilon_{2})\} =
\delta(\varepsilon_{1}-\varepsilon_{2}) H(\vec{x})$ where $
H(\vec{x};\varepsilon) = \langle \vec{x}| \delta(\varepsilon-
\mathcal{H}) |\vec{x} \rangle$ is the spectral Husimi function.
Consequently, the projected fields obeys
\begin{equation}
  \big\{ \psi_{\pm}(\vec{x}; \varepsilon_{1}), \psi_{\pm}^{\dagger}
  (\vec{x}; \varepsilon_{2}) \big\} 
= \frac{(\pm i/2\pi) H(\vec{x}),}{\varepsilon_{1}-\varepsilon_{2} \pm
  i\eta} 
\label{eq:Husimi-comm2}
\end{equation}
By using the above, we can finally write the current algebra Eq.
(\ref{eq:current-comm1}) as
\begin{equation}
 \left[ j(\vec{x};\varepsilon_{1}), j(\vec{x}; \varepsilon_{2})
\right]  = \frac{i}{2\pi} H(\vec{x};\varepsilon_{1}) H(\vec{x};\varepsilon_{2})
\delta'(\varepsilon_{1}-\varepsilon_{2}).
\label{eq:current-Husimi1}  
\end{equation}
This reveals clearly that
$j(\vec{x},\varepsilon)/H(\vec{x},\varepsilon)$ satisfies the Abelian
Kac-Moody algebra \emph{exactly}.  Now we can complete bosonization
(dual field formulation) at each $\vec{x}$ by introducing chiral boson
fields $\varphi(\vec{x};\varepsilon)$
\begin{equation}
  [\varphi(\vec{x};\varepsilon_{1}), \varphi(\vec{x};\varepsilon_{2})
  ] = - i\pi \sgn (\varepsilon_{1}-\varepsilon_{2}),
\end{equation}
by rewriting the current as
\begin{equation}
  j(\vec{x};\varepsilon) = H(\vec{x};\varepsilon) \, 
  \partial_{\varepsilon} \varphi(\vec{x};\varepsilon).
\label{eq:effective1}
\end{equation}
Note that the dual field $\varphi$ is meaningful only when the
``amplitude'' $H(\vec{x};\varepsilon)$ does not vanish.  They exist
only near energy levels and  the mode-locking is
fulfilled in this sense.

In passing, it is worth pointing out the difference between the Husimi
and the Wigner-Weyl representations.  Within the latter, it appears
possible to derive an algebra similar to Eq.
(\ref{eq:current-Husimi1}) \emph{approximately} by a semiclassical
expansion where the spectral Wigner function shows up instead of the
spectral Husimi function.  However, a wild oscillation of the Wigner
function makes such a semiclassical expansion difficult to justify in
general.  In contrast, the Husimi representation helps us identify the
symmetry \emph{exactly}.  Moreover since the Husimi function can be
viewed as a Gaussian smoothing of the Wigner function, it has a
well-defined semiclassical limit as a coarse-grained classical
dynamics \cite{Takahashi89,Almeida98}.  It is noted that from a
classical point of view, the energy level itself is taken as a caustic
because of a vanishing chord \cite{Almeida98}, so that the present
approach effectively extracts information of the time scale longer
than the Ehrenfest time.


We are so far concerned only with the symmetry of a single energy
shell $\varepsilon$ as appears in $\det(\varepsilon -\mathcal{H})$.
Since the $n$-point spectral correlation can be generated from the
$n$-fold ratio of the determinant correlator $\prod_{i=1}^{n}
\det(\varepsilon_{fi}-\mathcal{H})/\det(\varepsilon_{bi}-\mathcal{H})$,
we need to take account of additional degrees of freedom: the graded
(boson-fermion) symmetry and the internal symmetry among $n$ energy
shells.  As a result, the relevant symmetry is enlarged to a general
linear Lie superalgebra $\mathfrak{g}=\mathfrak{gl}(n|n)$ in the
simplest case (the unitary class), for which the preceding treatment
can be extended with minimal modification.  Explicitly, we introduce
the superbracket $\llbracket \cdot, \cdot \rrbracket$ and write the
commutation/anti-commutation relations for bose/fermion fields as
$\llbracket \psi_{\alpha}, \psi_{\beta}^{\dagger} \rrbracket =
\delta_{\alpha\beta}$.  By introducing the wave-packet (super-spinor)
field operator $\psi(\vec{x};\varepsilon)$, the normal-ordered current
operators on the Husimi representation are defined by
\begin{equation}
  j_{a}(\vec{x},;\varepsilon) = \;
  :\psi^{\dagger}(\vec{x};\varepsilon) X_{a} \psi(\vec{x};\varepsilon)
  : 
\end{equation}
where $X_{a} \in i\mathfrak{g}$ is an Hermitian element of a given Lie
superalgebra $\mathfrak{g}$ obeying $\llbracket X_{a}, X_{b}
\rrbracket = i f_{ab}^{c} X_{c}$. 
One  can evaluate the current commutator on the Husimi representation
as before to find
\begin{eqnarray}
&& \left\llbracket j_{a}(\vec{x}; \varepsilon_{1}), j_{b}(\vec{x};
  \varepsilon_{2}) \right\rrbracket = i f_{ab}^{c}\,
  \delta(\varepsilon_{1}-\varepsilon_{2}) H(\vec{x}; \varepsilon_{2})
  j(\vec{x}; \varepsilon_{2}) \nonumber \\
&& \qquad + \frac{i\, \kappa_{ab}}{2\pi} H(\vec{x}; \varepsilon_{1})
  H(\vec{x}; \varepsilon_{2}) \delta'(\varepsilon_{1}-\varepsilon_{2}),
\label{eq:comm-Husimi2}
\end{eqnarray}
where $\kappa_{ab} = \STr[X_{a} X_{b}]$.  This is the main result of
the paper.  It shows clearly that $j_{a}(\vec{x};
\varepsilon)/H(\vec{x};\varepsilon)$ satisfies the Kac-Moody algebra
of a corresponding Lie superalgebra.  Hence the effective field theory
is described by the (chiral) Wess-Zumino-Novikov-Witten model defined
on a corresponding Lie supergroup with the current operator
\begin{equation}
  j(\vec{x}; \varepsilon) = H(\vec{x}; \varepsilon) \;
  g^{-1}\partial_{\varepsilon} g(\vec{x};
  \varepsilon).  
\label{eq:effective2}
\end{equation}


By recognizing the convoluted function $(\varepsilon - \varepsilon'\pm
i\eta)^{-1}$ in Eq. (\ref{eq:projectedfield}) is a Fourier transform
of the step function, the projection onto $(\pm)$ may be regarded as
the decomposition into the retarded $(R)$ and advanced $(A)$
components.  We can make the correspondence explicit by writing
\begin{equation}
  \psi_{+}(\vec{x};\varepsilon) =  \left( \begin{array}{c}
  b_{R} \\ f_{R}  \end{array}\right) ; \quad  
  \psi_{-}(\vec{x}; \varepsilon) = 
  \left(\begin{array}{c} - b_{A}^{\dagger} \\  f_{A}^{\dagger}
  \end{array} \right), 
\end{equation}
where $b_{R,A}$ ($f_{R,A}$) are taken as bosonic (fermionic) fields to
generate the retarded/advanced Green functions.  The minus sign of
$b_{A}^{\dagger}$ is mandatory to retain the commutation relations.
The condition of the vacuum state becomes $b_{R}|0\rangle =
f_{R}|0\rangle = b_{A}|0\rangle = f_{A}|0\rangle = 0$ so that the
definition of the normal ordering coincides with the standard
definition in the $bf$-fields.
From here, one can construct the color-flavor transformation by using
the coherent states of $bf$-fields at each $\vec{x}$, as is given in
Appendix of \cite{Zirnbauer96b}. Hence the supermatrix NL-$\sigma$
model.  The only modification that is crucial for the mode-locking
problem is the presence of the spectral Husimi function
$H(\vec{x};\varepsilon)$ instead of the average DOS as a result of the
commutation relation Eq. (\ref{eq:Husimi-comm2}).  In this way, the
supermatrix NL-$\sigma$ model with the exact DOS, $h^{-d}\int
H(\vec{x};\varepsilon) d\vec{x}$, can be derived in a ballistic
system.  The zero-dimensional approximation leads to the universal
Wigner-Dyson correlation, with nonuniversal deviation from the
\emph{coarse-grained} semiclassical dynamics of the Husimi function.
Further additional coarse-graining or semiclassical approximation
gives a smoothing of DOS, which is believed to correspond to the
situation argued in
\cite{Muzykantskii95Efetov03,Andreev96,Altland99b,Zirnbauer99b,Efetov04}.

In the present work, we are concerned with the simplest case that the
phase space $\vec{x}$ has no additional symmetry (the unitary class).
It is not the case in the time-reversal symmetric system, where both
the symmetry of equivalent vacua and the corresponding algebra need to
be enlarged by the time-reversal operation, {\it i.e.}, mixing between
an original field and its Hermitian conjugate counterpart.  The
relevant symmetry of the $n$-point level correlation in this case is
identified as an orthosymplectic algebra $\mathfrak{osp}(2n|2n)$,
which coincides with the symmetry of the Bogoliubov transformation of
the enlarged space.  Any additional discrete/continuous symmetry will
be accommodated similarly.

It is stressed that fields appearing in the effective field
description result essentially from the phase arbitrariness of each
eigen wavefunction, {\it i.e.}, $U(1)$ symmetry at each level (or the
Berry phase).  By putting quantum chaos aside, it is worth thinking of
the implication of the present construction in interacting electrons,
where we no longer have $U(1)$ symmetry to each level but only one
global $U(1)$ phase at the Fermi energy.  This suggests that the
present construction may still be meaningful at the Fermi energy and
should be closely related to the Luther-Haldane bosonization for
interacting electrons \cite{Haldane94} (by including the degeneracy
properly).  However, an explicit construction toward it is open at
present.

In conclusion, we have examined the effective field theory of a
quantum chaotic billiard from the perspective of quantum anomalies,
and the theory is shown to be endowed with a symmetry of the Kac-Moody
algebra exactly.  This allows one to formulate the supermatrix
NL-$\sigma$ model without introducing any additional coarse-graining
nor stochasticity.  It is also found that the use of the Husimi
representation is indispensable to identifying the correct symmetry,
and the spectral Husimi function acts as the amplitude of the
effective fields.


I thank A. Tanaka for helpful comments on bosonization methods.  

%

\end{document}